\documentclass[aps,prl,twocolumn,groupedaddress]{revtex4-1}

\usepackage{graphicx}
\usepackage{mathtools}
\usepackage{amsmath,amssymb}
\usepackage[caption=false]{subfig}

\begin{document}


\title{Streamwise localization of traveling wave solutions in channel flow}


\author{Joshua Barnett}
\author{Daniel R. Gurevich}
\author{Roman O. Grigoriev}
\affiliation{School of Physics, Georgia Institute of Technology, Atlanta, Georgia 30332-0430, USA}


\date{\today}

\begin{abstract}
Channel flow of an incompressible fluid at Reynolds numbers above 2400 possesses a number of different spatially localized solutions that approach laminar flow far upstream and downstream. We use one such relative time-periodic solution, which corresponds to a spatially localized version of a Tollmien-Schlichting wave, to illustrate how the upstream and downstream asymptotics can be computed analytically. In particular, we show that for these spanwise uniform states the asymptotics predict the exponential localization that has been observed for numerically computed solutions of several canonical shear flows but never properly understood theoretically.
\end{abstract}


\maketitle

Due to the development of new computational techniques as well as an increase in computing power, our understanding of fluid flows in the transitional and weakly turbulent regime has improved rather dramatically in the past couple of decades. 
What started as a numerical exploration of minimal flow units with unphysical (e.g., spatially periodic) boundary conditions \cite{nagata1990,kawahara2001,waleffe2001} has eventually yielded a large number of both stable and unstable nontrivial solutions to the Navier-Stokes equation featuring simple temporal dynamics. 
These solutions have produced significant insight into coherent structures \cite{hussain1983,sirovich1987} that have been routinely observed in experiments and into the self-sustaining physical processes maintaining turbulence in wall-bounded shear flows \cite{waleffe1997,wang2007}.

In order to better connect numerical results with experiments, computational domains have steadily increased in both the spanwise and streamwise directions. 
More recent studies discovered that many spatially-periodic solutions start to localize in the spanwise direction \cite{schneider2010,gibson2014}, in the streamwise direction \cite{mellibovsky2015}, or both \cite{brand2014,zammert2014}, as the domain size increases. 
In both cases localization results from subharmonic instability. 
In particular, spanwise localization generates a discrete set of solutions with different width due to the snaking mechanism \cite{burke2007}. The mechanism that controls streamwise localization is however not entirely clear.

Spatial localization plays a crucial role both in extending results obtained on relatively small computational domains to arbitrarily large physical domains and in understanding how different regions of weakly turbulent flows interact with each other. 
For instance, while temporal aspects of intermittency in weakly turbulent flows were understood with the help of dynamical systems theory \cite{pomeau1980} a long time ago, the spatial organization of intermittent flows (e.g., the formation of turbulent bands separated by laminar regions) \cite{wygnanski1973,barkley2005,meseguer2009} is still an open problem, despite some recent advances \cite{chantry2016}. 

Quite a few of the spatially localized solutions computed numerically have been found to exhibit an exponential localization. 
Although there is some theoretical support for exponential localization in the streamwise direction \cite{gibson2014}, there is little understanding of the scaling of the upstream and downstream tails of streamwise-localized solutions, the spatial modulation of these tails, or the mechanism that controls the drift speed of streamwise-localized solutions. 
The aim of this communication is to show that streamwise asymptotics of several dynamically important classes of spatially localized solutions can be described very accurately by using particular solutions of the Orr-Sommerfeld equation.

We will focus on the flow of an incompressible fluid through a channel with parallel planar walls. It is described by the Navier-Stokes equation which, after nondimensionalization by the channel width, takes the form 
\begin{align}\label{eq:nse}
\partial_t{\bf v}+{\bf v}\cdot\nabla{\bf v}=-\nabla p+Re^{-1}\nabla^2{\bf v},
\end{align}
where $p$ is the nondimensional pressure, $x$ is the streamwise, $y$ is the wall-normal, and $z$ is the spanwise direction. The laminar solution satisfying the no-slip boundary conditions for the velocity ${\bf v}$ at the walls of the channel ($y=\pm1$) is known as the plane Poiseuille flow: ${\bf v}_0=[U(y),0,0]$, where $U=1-y^2$. Writing ${\bf v}=[U+u,v,w]$ and linearizing \eqref{eq:nse} about the laminar flow profile gives \cite{schmid2012} the Orr-Sommerfeld equation
\begin{align}\label{eq:ose}
\left(\partial_t+U\partial_x-Re^{-1}\nabla^2\right)\nabla^2v-U''\partial_xv=0
\end{align}
for the wall-normal velocity $v$ and the Squires equation 
\begin{align}\label{eq:sqe}
\left(\partial_t+U\partial_x-Re^{-1}\nabla^2\right)\eta+U''\partial_zv=0
\end{align}
for the wall-normal vorticity $\eta=\partial_zu-\partial_xw$. It is customary to look for solutions to \eqref{eq:ose} and \eqref{eq:sqe} in the form $v=\hat{v}(y)e^{i\alpha x+i\beta z+\lambda t}$ and $\eta=\hat{\eta}(y)e^{i\alpha x+i\beta z+\lambda t}$, which describe three-dimensional disturbances. In this case $\partial_t[\cdot]=\lambda[\cdot]$, $\nabla^2[\cdot]=(\partial_y^2-\alpha^2-\beta^2)[\cdot]$, yielding one-dimensional boundary value problems for $\hat{v}(y)$ and $\hat{\eta}(y)$. 

\begin{figure*}
\includegraphics[width=\textwidth]{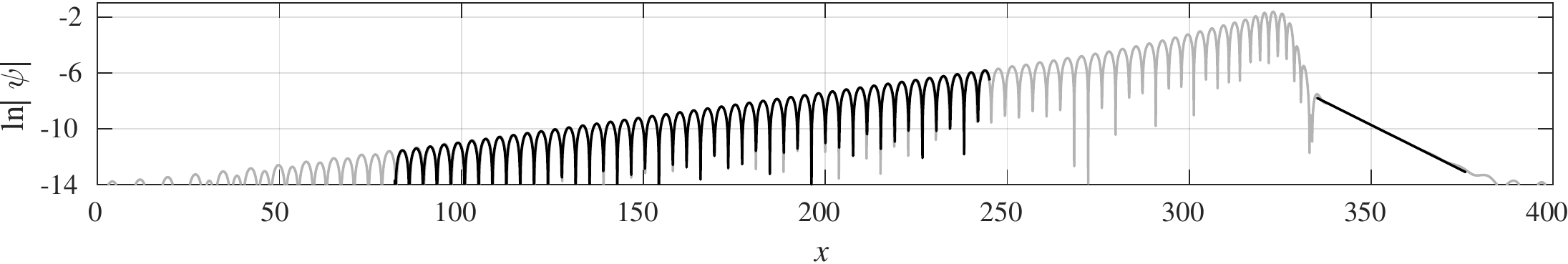}
\caption{\label{fig:lognorm}A snapshot of the magnitude of the stream function $\psi$ describing the MTSW at $Re=3802$ on the axis $y=0$ of the channel. Numerical solution is shown in gray, the fits based on the solutions to the Orr-Sommerfeld equation for the leading and trailing tail are shown in black. Numerical accuracy of the solution corresponds to $\ln|\psi|\approx-14$.}
\end{figure*}

The Orr-Sommerfeld equation \eqref{eq:ose} can be solved independently; its solution relates the streamwise wavenumber $\alpha$ and spanwise wavenumber $\beta$ of an infinitesimal disturbance, its transverse profile $\hat{v}(y)$, and the eigenvalue (stability exponent) $\lambda=\sigma+i\omega$. The spectrum of the corresponding boundary value problem is discrete, with infinitely many solutions; we will focus on the one which corresponds to the eigenvalue with the largest real part. Furthermore, since we are primarily interested in streamwise localization, we will only consider two-dimensional (2D) disturbances for which $\beta=0$, $\eta=0$, and both wall-normal and streamwise velocity components can be written in terms of a stream function: $u=\partial_y\psi$, $v=-\partial_x\psi$, where $\psi(x,y,t)=\phi(y)e^{i\alpha x+\lambda t}$

Using the Orr-Sommerfeld equation, Orszag \cite{orszag1971} has shown that at $Re_c=5772.22$ the laminar solution becomes unstable towards a spatially periodic modulation with wavenumber $\alpha_c=1.02056$, and above $Re_c$ the flow becomes turbulent. However, even below $Re_c$, multiple solutions of \eqref{eq:nse} have been found, both stable and unstable. One example is nonlinearly saturated 2D Tollmien-Schlichting waves (TSW), which have a spatially uniform envelope. Mellibovsky and Meseguer \cite{mellibovsky2015} have recently found a family of 2D localized solutions, termed modulated Tollmien-Schlichting waves (MTSW), related to TSW via a spatial subharmonic instability \cite{drissi1999}. TSW correspond to relative equilibria and MTSW to relative periodic orbits: the former become stationary and the latter, temporally periodic in a reference frame moving with some velocity $c>0$ relative to the walls of the channel. For both types of solutions $c$ is the group velocity. The phase velocity is also equal to $c$ for TSW but is different from $c$ for MTSW.

\begin{figure*}
\includegraphics[width=0.9\textwidth]{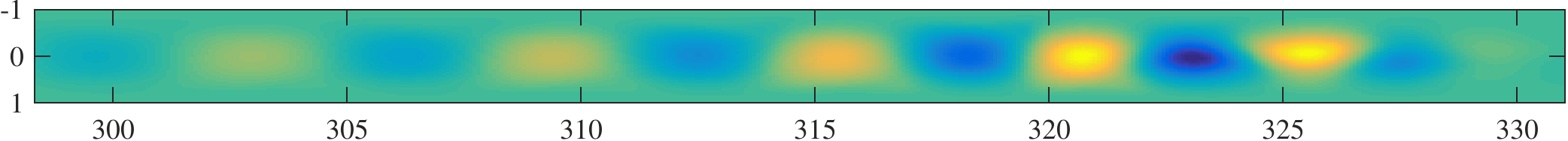}\\
(a)\\\vspace{1mm}
\includegraphics[width=0.44\textwidth]{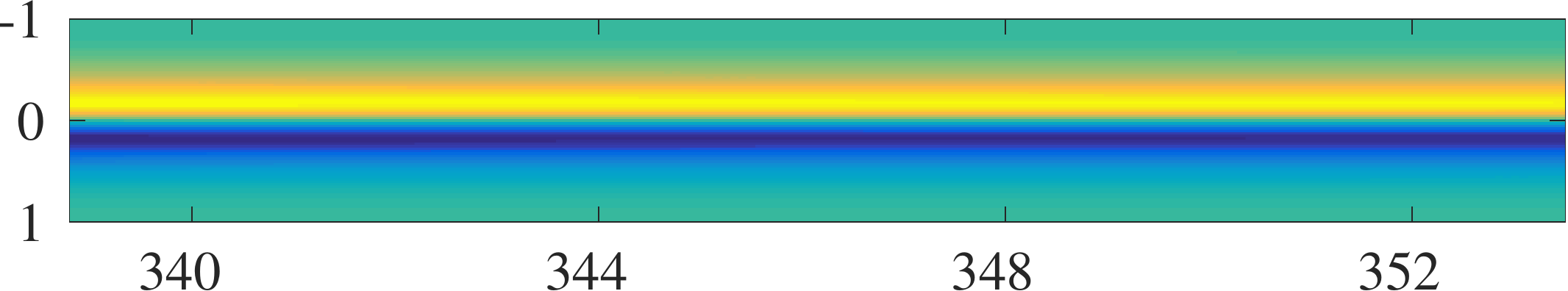}\hspace{3mm}
\includegraphics[width=0.44\textwidth]{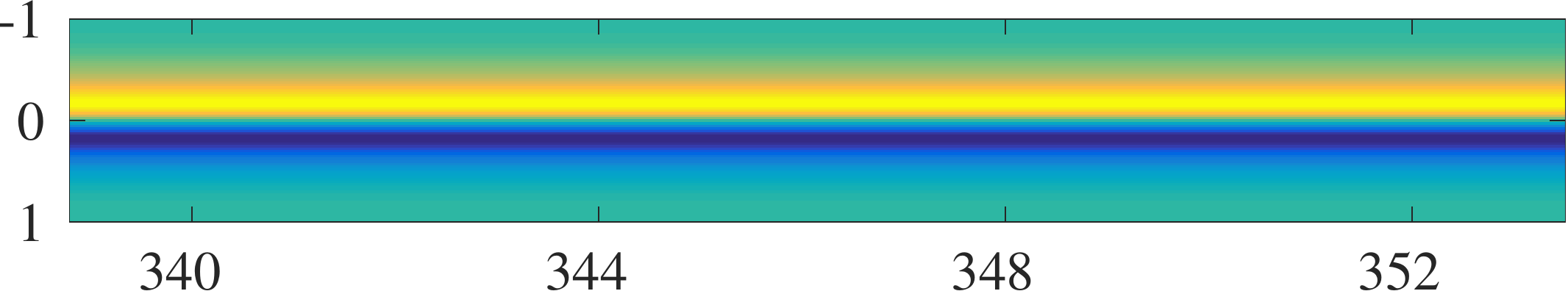}\\
(b)\\\vspace{1mm}
\includegraphics[width=0.44\textwidth]{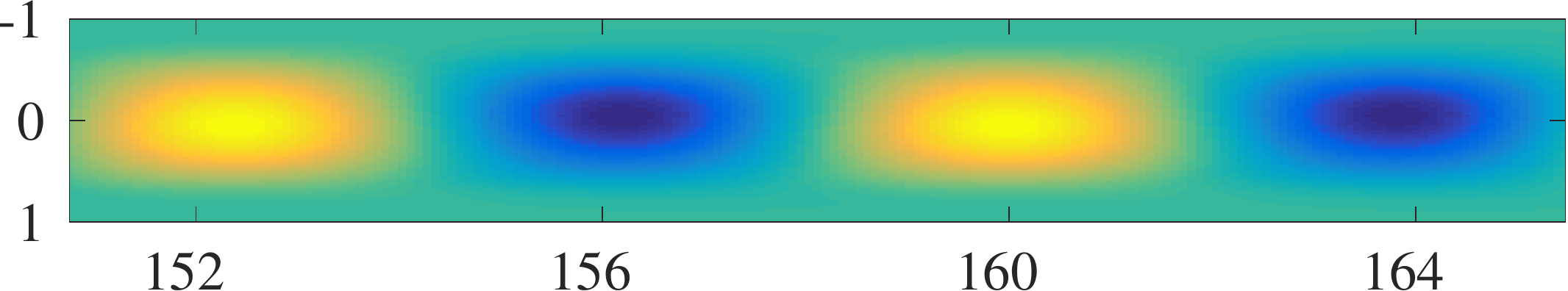}\hspace{3mm}
\includegraphics[width=0.44\textwidth]{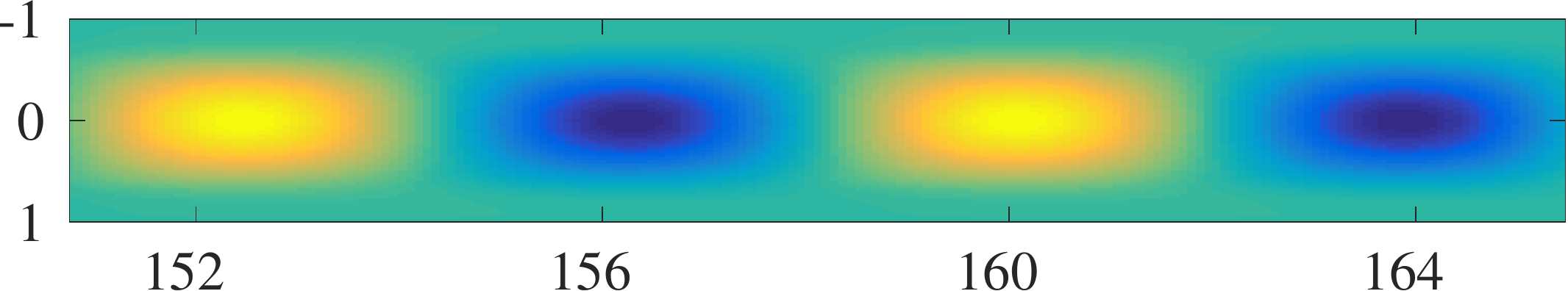}\\
(c)\\
\caption{\label{fig:psi}The stream function $\psi$ describing the MTSW at $Re=3802$. (a) The core region. (b) The scaled stream function $e^{s_1x}\psi$ for a segment of the leading edge: numerical solution (left) and analytical solution (right). (c) The scaled stream function $e^{s_2x}\psi$ for a segment of the trailing edge: numerical solution (left) and analytical solution (right). The horizontal and vertical axes correspond to $x$ and $y$, respectively, with the $x$ coordinate corresponding to that in Fig. \ref{fig:lognorm}. See Fig. \ref{fig:connection}(b) for the colormap.}
\end{figure*}

We have computed MTSW for several different $Re$ on a domain of length $L_x=400$ (with periodic boundary conditions in the $x$ direction) using the package {\it Channelflow} \cite{channelflow}. The stream function associated with one such solution at $Re=3802$ is shown in Figs. \ref{fig:lognorm} and \ref{fig:psi}(a). A distinguishing feature of all MTSW is that their localization is exponential, both in the upstream and in the downstream direction, with the solution approaching a laminar profile ${\bf v}_0$ for $x\to\pm\infty$. This localization can also be understood using the Orr-Sommerfeld equation.

Unlike the stability analysis, we should look at disturbances with a streamwise wavenumber that is complex, $\alpha=q+is$, where the real part $q$ describes the spatial modulation and the imaginary part $s$, the spatial attenuation of the tails of a localized solution. Let $\xi=x-ct$ describe the streamwise coordinate in the reference frame moving with speed $c$ (i.e., the group velocity of the MTSW), such that, for $\xi\to\pm\infty$, the tails of the solution can be written in the form $\psi(\xi,y,t)=\phi(y)e^{iq\xi-s\xi}e^{\sigma't}e^{i\omega't}$. Since the MTSW is temporally periodic in this reference frame, we should have
\begin{align}\label{eq:match}
\sigma'\equiv\sigma(q,s)-cs&=0,\nonumber\\
\omega'\equiv\omega(q,s)+cq&=\frac{2\pi}{T}n,
\end{align}
where $T$ is the temporal period of the MTSW and $n$ is an integer. The system of equations \eqref{eq:match} can be solved with the help of the Matlab package {\it chebfun} \cite{chebfun} and possesses several solutions for the relevant range of Reynolds numbers. Those with $s>0$ describe the downstream/leading tail ($\psi\to0$ for $\xi\to\infty$) and those with $s<0$, the upstream/trailing tail ($\psi\to0$ for $\xi\to-\infty$). Note that the leading mode of \eqref{eq:ose} may not satisfy conditions \eqref{eq:match} for either positive or negative $s$ for any $q$ and $n$. In this case the next leading mode has to be considered, etc.

In particular, for $Re=3802$ we find three solutions ${\bf a}=(q,s,n)$: ${\bf a}_1=(0,0.130,0)$, ${\bf a}_2=(0.826,-0.0354,1)$, and ${\bf a}_3=(0,-0.00642,0)$.  The first two accurately describe, respectively, the leading and the trailing tail of the corresponding MTSW, as Fig. \ref{fig:lognorm} illustrates. The best fit values found using the fully nonlinear numerical solution (${\bf a}_1^{\mathrm{num}}=(0,0.129,0)$ for the leading tail and ${\bf a}_2^{\mathrm{num}}=(0.826,-0.355,0)$ for the trailing tail) are in excellent agreement with predictions based on linearization.

We can further confirm the accuracy of the analytical approach by comparing the spatial profiles of the stream function computed using the Navier-Stokes equation \eqref{eq:nse} and its linearization (i.e., the Orr-Sommerfeld equation). To amplify the attenuated regions of each tail, we multiply each stream function by the corresponding exponential factor, i.e., $e^{s_1x}\psi$ for the leading tail and $e^{s_2x}\psi$ for the trailing tail. The results are compared in Fig. \ref{fig:psi}(b) for the leading tail and in Fig. \ref{fig:psi}(c) for the trailing tail. In both cases we again find excellent agreement.

The structure of the tails has an interesting physical interpretation. In the stationary reference frame, the leading tail describes a disturbance $\psi\propto\phi_1(y)e^{i\alpha_1x+\lambda_1t}$ about the laminar flow profile that is unstable ($\sigma>0$), vanishes at $x\to\infty$ ($s>0$), and has no spatial ($q=0$) or temporal ($\omega=0$) modulation. The transverse profile of this disturbance (cf. Fig. \ref{fig:psi}(b)) corresponds to a monotonic increase in the shear rate of the flow in the central region $|y|<0.45$ of the channel. This increase in the shear can be thought of as promoting the Kelvin-Helmholtz instability, which, unlike the initial disturbance, has an oscillatory character. The core of the MTSW (which we define as the region shown in Fig. \ref{fig:psi}(a), where $|\psi|$ and the velocity perturbation are the largest) can therefore be thought of as a result of nonlinear saturation in the spatial and temporal modulation.

\begin{figure}
\centering
\includegraphics[width=0.56\columnwidth]{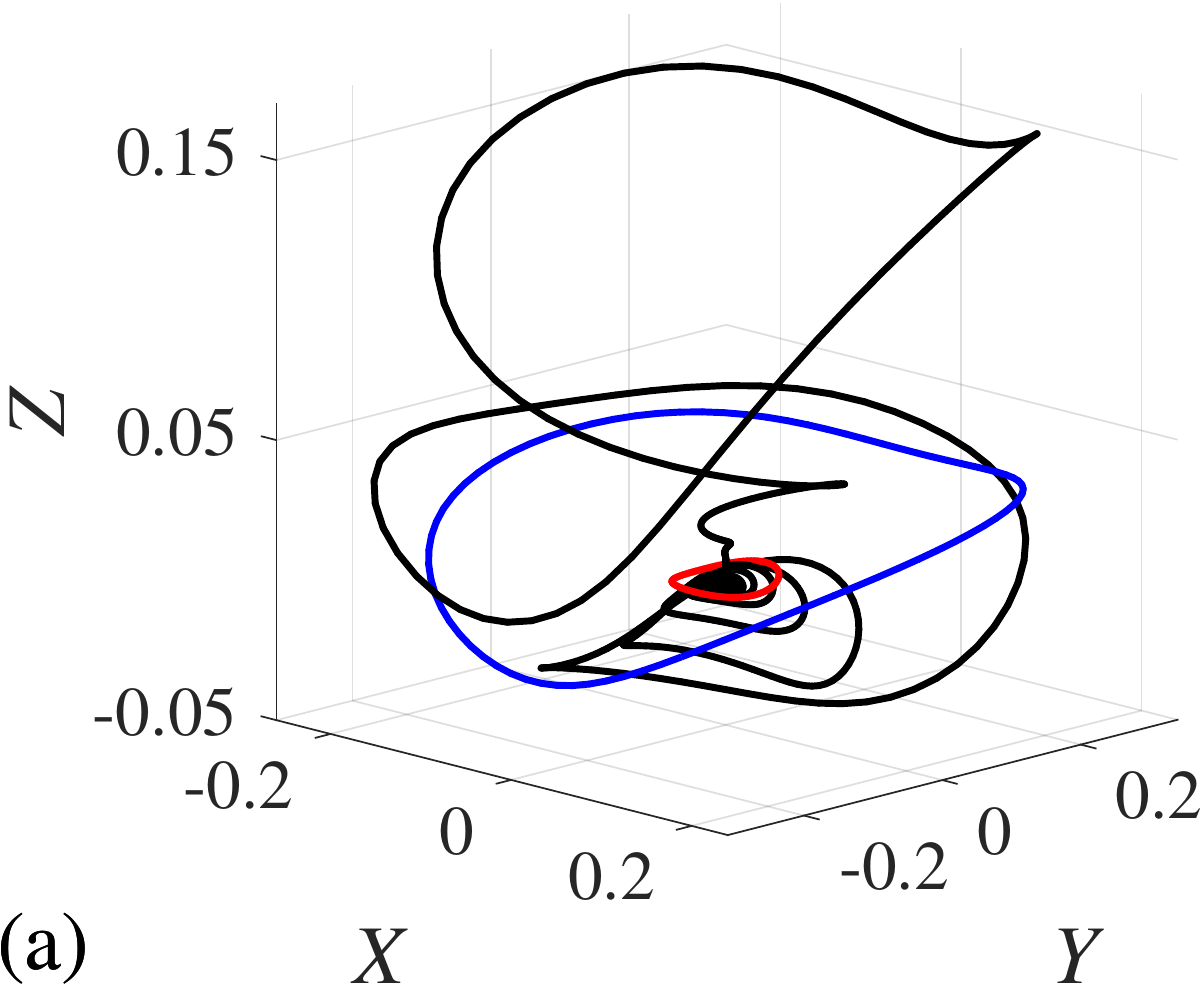}
\hspace{2mm}
\includegraphics[width=0.38\columnwidth]{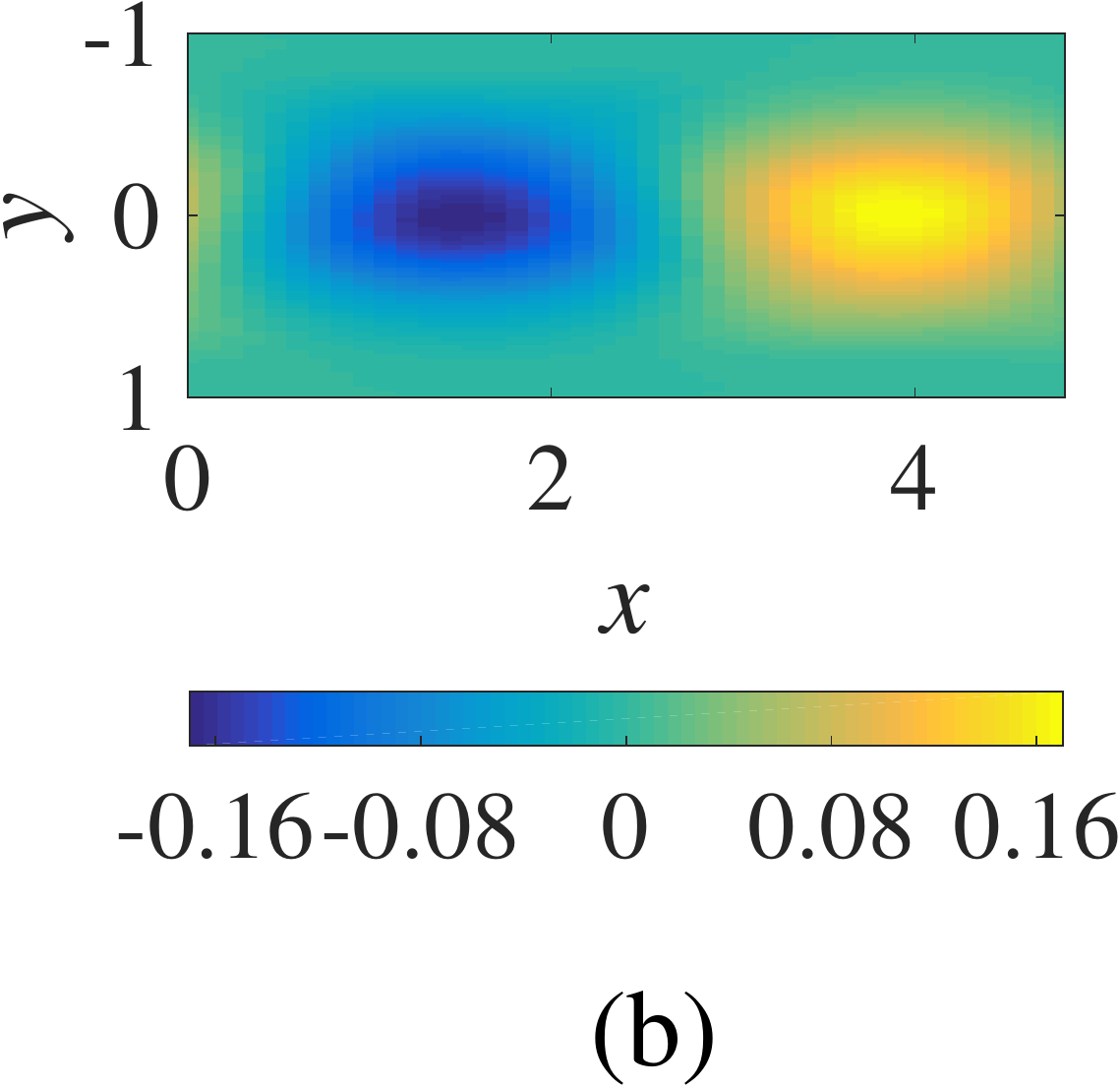}
\caption{\label{fig:connection} (a) The homoclinic orbit. MTSW and the corresponding upper and lower branch TSW \cite{mellibovsky2015} are shown as black, blue, and red curves, respectively. The coordinates are $X=v$, $Y=\partial_x v$, and $Z=u$ at $y=0$ and $t$-fixed, with the origin corresponding to laminar flow. (b) The stream function $\psi$ for the upper branch TSW.}
\end{figure}

Eventually the perturbation starts to decay back to the laminar state. This decay is controlled by the two solutions of the Orr-Sommerfeld equation with $s<0$ and $\sigma<0$. Hence, in general, one might expect to see a linear combination of these two solutions
\begin{align}\label{eq:trail}
\psi=A_2\phi_2(y)e^{i\alpha_2x+\lambda_2t}+A_3\phi_3(y)e^{i\alpha_3x+\lambda_3t}
\end{align}
with the coefficients $A_2$ and $A_3$ determined by the boundary conditions in the region where the trailing tail is matched to the core region. Since the temporal dynamics in the core region are dominated by а nearly harmonic modulation with frequency $2\pi/T$, it is natural to expect the mode with $n=1$ to dominate over the mode with $n=0$ (i.e., $|A_2|\gg|A_3|$), so the second term in \eqref{eq:trail} can be neglected. This is indeed what we find: the first term alone describes both the streamwise profile of the trailing tail of the numerical solution (cf. Fig. \ref{fig:lognorm}) and its wall-normal profile (cf. Fig. \ref{fig:psi}(c)).

If we fix $t$ and vary $x$ (or fix $x$ and vary $t$), the MTSW becomes a homoclinic orbit that starts and ends at the laminar state (cf. Fig. \ref{fig:connection}(a)). The leading tail (aligned along the $Z$ axis in this projection) describes the transition from laminar flow to the neighborhood of a corresponding upper branch TSW shown in Fig. \ref{fig:connection}(b). Since this TSW is itself unstable on a long domain \cite{drissi1999}, the flow eventually leaves its neighborhood and spirals back towards the laminar solution, bypassing the lower branch TSW. This final piece of the orbit corresponds to the trailing tail.

The group velocity $c$ has been computed simultaneously with MTSWs using a matrix-free Newton-Krylov method \cite{waleffe1998}. It is natural to ask whether $c$ can instead be found from linearization. Front propagation theory predicts that, for a pulled front, the front speed and streamwise wavenumber are uniquely defined by a solution to the following system of equations \cite{vansaarlos1989}:
\begin{align}\label{eq:front}
c=i\frac{\partial\lambda}{\partial\alpha}=\frac{\mathrm{Re}(\lambda)}{\mathrm{Im}(\alpha)},
\end{align}
which, in particular, requires $\partial\sigma/\partial s=\sigma/s$. For channel flow, this equation has no solutions for $q=0$ in the range of $Re$ we explored, so the speed of the MTSW is not selected by a linear mechanism (i.e., the leading tail corresponds to a pushed front). Therefore, the speed of MTSW is controlled by the core region, rather than the leading tail, and hence the drift speed should be different for different localized solutions. Furthermore, since the first of the equalities in \eqref{eq:front} does not hold, neither does the conventional relationship $c=-\partial\omega/\partial q$ for the group velocity. Indeed, the system \eqref{eq:match} has a discrete set of solutions, so the variation of the temporal frequency with spatial wavenumber is meaningless.

\begin{figure}
\centering
\includegraphics[width=0.91\columnwidth]{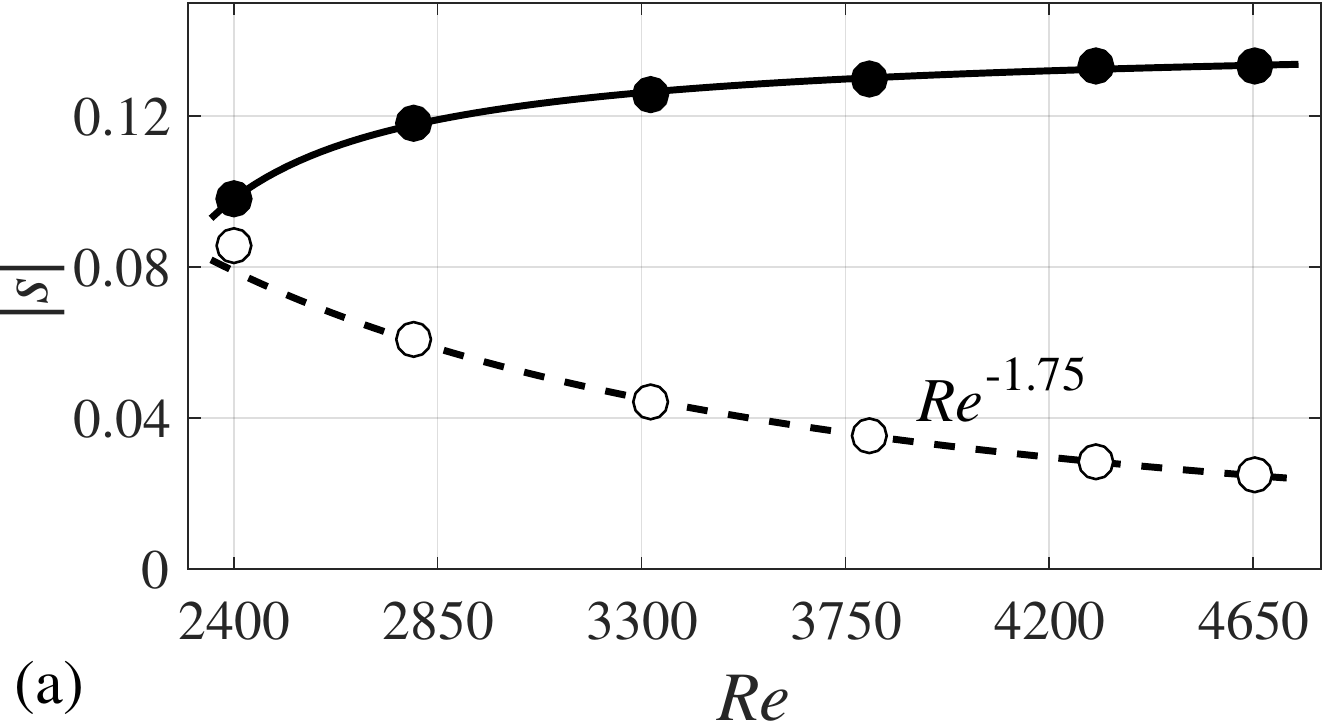}
\hspace{\columnwidth}\\\vspace{1mm}\hspace{1mm}
\includegraphics[width=0.9\columnwidth]{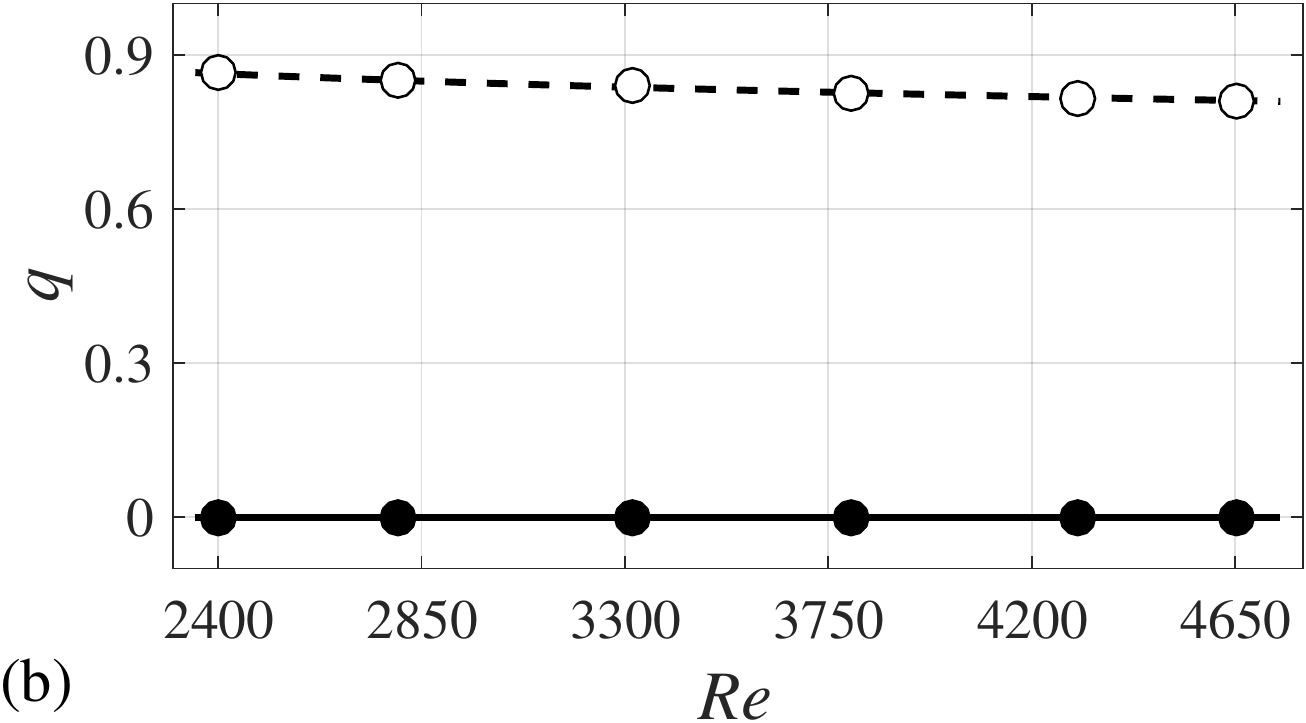}
\caption{\label{fig:alpha}The spatial decay rate $|s|$ (a) and the wavenumber $q$ (b) of the tails of MTSW. The data for the leading (trailing) tail are shown as filled (open) symbols. MTSW is stable for $Re\le 3802$ and unstable for $Re\ge 4300$.}
\end{figure}

It has been noticed previously that the leading and the trailing tail of streamwise-localized solutions of shear flows decay spatially at different rates. In particular, for doubly-localized solutions of channel flow, it was found that the leading tail decays at a rate that is approximately constant, while the trailing tail decays at a rate $|s|\propto Re^{-1}$ \cite{zammert2014}. We find that the general trends are similar for MTSW which lack spanwise localization: the decay rate $s_1$ for the leading tail increases weakly with $Re$ (cf. Fig. \ref{fig:alpha}(a)), while the decay rate for the trailing tail decreases as a power law $|s_2|\propto Re^\gamma$ with the exponent $\gamma=-1.75$. In both cases the localization is the strongest at smaller $Re$ and solutions gradually de-localize as $Re$ increases. The general agreement suggests that the description presented here should apply in equal measure to solutions which have both streamwise and spanwise localization, although the specific results would depend on the properties of a particular solution. For instance, for MTSW, the leading tail does not have spatial modulation and the trailing tail does (cf. Fig. \ref{fig:alpha}(b)), while for doubly-localized solutions the opposite is true. 

To conclude, we have demonstrated that the asymptotics of streamwise-localized 2D modulated Tollmien-Schlichting waves in channel flow are well-described by solutions to the Orr-Sommerfeld equation with a complex streamwise wavenumber. This approach is capable of describing both the streamwise and spanwise asymptotics of localized solutions in any shear flow, provided that they are described by either relative equilibria or relative periodic orbits, regardless of their stability. Spanwise localization can be described using solutions of the Orr-Sommerfeld equation with real $\alpha$ and complex $\beta\ne 0$ as shown previously for relative equilibria in plane Couette flow \cite{gibson2014}. For doubly localized solutions, all components of the velocity are nonzero \cite{zammert2014}, so, in addition to the Orr-Sommerfeld equation \eqref{eq:ose}, one also has to solve the Squire equation \eqref{eq:sqe} for complex $\alpha$ and $\beta$. The streamwise and spanwise velocity components can then be computed by solving $(\partial_x^2+\partial_z^2)u=\partial_z\eta-\partial_x\partial_yv$ and $(\partial_x^2+\partial_z^2)w=-\partial_x\eta-\partial_y\partial_yv$.

Due to its linearity, the Orr-Sommerfeld equation is dramatically easier to study analytically compared with the nonlinear Navier-Stokes equation. As an example, the approximate equation describing streamwise localization of a particular absolute equilibrium of plane Couette flow derived by retaining only the dominant terms in the Navier-Stokes equation \cite{brand2014} corresponds to a time-independent Orr-Sommerfeld equation. Until recently, the fully nonlinear solutions of Navier-Stokes could be computed and studied only numerically. The results presented here and in related studies \cite{gibson2014,zammert2014,brand2014} give us hope that we can understand some of their properties analytically. In particular, we found that the group velocity of localized solutions is controlled by a nonlinear mechanism and hence is not universal, which has important implications for the structure and dynamics of turbulent bands/spots/puffs in various intermittent flows.

\begin{acknowledgments}
This work was supported in part by the National Science Foundation under Grant No. CMMI-1234436. We are grateful to F. Mellibovsky and A. Meseguer for sharing some of the localized solutions they have computed and to J. Gibson for sharing his numerical code {\it Channelflow} and for helping us set up the simulations. We would also like to thank an anonymous referee for pointing out that, after the original version of this manuscript had been submitted for publication, a related paper \cite{zammert2016} has appeared that considers doubly localized solutions of channel flow.

\end{acknowledgments}

\bibliography{channelflow.bib}

\end{document}